%% file: template.tex
\documentclass{Interspeech}

\interspeechcameraready

\title{Prosody-Adaptable Audio Codecs for Zero-Shot Voice Conversion via In-Context Learning}

\author[affiliation={1}*]{Junchuan}{Zhao}
\author[affiliation={1}*]{Xintong}{Wang}
\author[affiliation={1}]{Ye}{Wang}

\affiliation{School of Computing}{National University of Singapore}{Singapore}
\email{junchuan@u.nus.edu, xintong.wang@u.nus.edu, dcswangy@nus.edu.sg}

\usepackage{comment}
\usepackage{marginnote}

\begin{document}

\maketitle

\def\thefootnote{*}\footnotetext{These authors contributed equally to this research.}
\def\thefootnote{\arabic{footnote}}

\begin{abstract}\label{sec:abstract}
\input{sections/abstract}
\end{abstract}

\section{Introduction}\label{sec:intro}
\input{sections/intro.tex}


\section{Methodology}\label{sec:meth}
\input{sections/method}

\section{Experiments}\label{sec:exp}
\input{sections/results}

\section{Conclusion}\label{sec:concl}
\input{sections/concl}

\bibliographystyle{IEEEtran}
\bibliography{main}

\end{document}

%% file: sections/abstract.tex
Recent advances in discrete audio codecs have significantly improved speech representation modeling, while codec language models have enabled in-context learning for zero-shot speech synthesis. Inspired by this, we propose a voice conversion (VC) model within the VALLE-X framework, leveraging its strong in-context learning capabilities for speaker adaptation. To enhance prosody control, we introduce a prosody-aware audio codec encoder (PACE) module, which isolates and refines prosody from other sources, improving expressiveness and control. By integrating PACE into our VC model, we achieve greater flexibility in prosody manipulation while preserving speaker timbre. Experimental evaluation results demonstrate that our approach outperforms baseline VC systems in prosody preservation, timbre consistency, and overall naturalness, surpassing baseline VC systems.

%% file: sections/intro.tex
Voice conversion (VC) is an advanced speech processing technique that facilitates the transformation of one speaker's voice into another's while preserving the underlying content \cite{DBLP:conf/interspeech/LiuCKHL00M20, DBLP:journals/taslp/SismanYKL21}. By effectively separating linguistic content from speaker-specific features such as timbre and prosody, VC enables the synthesis of speech that preserves the original message while adopting the vocal characteristics of a target speaker. This technology has gained significant attention for its diverse applications, enhancing communication and user experience. VC is vital to preserve privacy through voice anonymization and empowers individuals with speech impairments to express their desired vocal identity. Its ability to manipulate voice attributes makes VC a transformative tool in speech synthesis and human-computer interaction, driving innovation and improving accessibility in various domains \cite{app13053100}. 

Recent advances in deep learning have transformed VC by improving the quality of synthesized voices, leading to more natural intonation, clearer articulation, and greater emotional expression for a more authentic auditory experience \cite{DBLP:conf/interspeech/LiZM21, DBLP:conf/icml/CasanovaWSJGP22, DBLP:conf/iclr/PopovVGSKW22, DBLP:conf/icassp/TangZWCX22, DBLP:conf/icassp/LeeKLYY20, DBLP:conf/icassp/LianZY22}. However, two major challenges still present opportunities for further advancement. The first is to achieve robust disentanglement between speech content and speech features, as well as between different features such as prosody and timbre, to enable more precise control over voice characteristics. The second challenge lies in improving zero-shot voice conversion, where the system must generate high-quality voice transformations for unseen speakers without requiring extensive training data. 

Most VC systems focus primarily on disentangling content from speaker-specific features \cite{DBLP:conf/interspeech/LiuHL19, DBLP:conf/interspeech/LuongT21, DBLP:conf/interspeech/ChouL19}, but disentangling within the speaker features themselves, such as separating prosody, timbre, and pitch, is equally critical. Achieving this finer level of separation is essential for enabling more expressive voice conversion and providing greater control over voice characteristics, allowing for nuanced transformations that capture the full range of human speech dynamics. There are several studies have concentrated on disentangling prosody and timbre for voice conversion \cite{DBLP:conf/iscslp/LianZWLT21, DBLP:conf/icassp/ZhaoLSWKTM22, DBLP:conf/ijcnn/WangB22}. \cite{DBLP:conf/icassp/ZhaoLSWKTM22} introduces an innovative method that employs a unit encoder, speaker verification, and a prosody encoder, enhanced by an adversarial content predictor. This approach effectively minimizes information overlap between prosody and content embedding, facilitating more distinct and controlled representations. \cite{DBLP:conf/ijcnn/WangB22} introduces a self-supervised method that learns disentangled pitch and volume representations from augmented speech, effectively capturing prosody styles and enhancing zero-shot voice conversion while mitigating prosody leakage. 

Zero-shot voice conversion remains a key challenge due to the need for systems to adapt to unseen speakers without prior training. Previous studies have employed speaker embeddings to capture timbre information from reference speakers, enabling systems to generalize to new voices without fine-tuning, thereby improving the adaptability of voice conversion technologies across diverse speaker identities \cite{DBLP:conf/icml/QianZCYH19, DBLP:conf/icassp/XiaoZL22}. However, a major limitation of this approach in both zero-shot voice conversion and text-to-speech (TTS) is the dependency on a robust, well-trained speaker encoder, which requires access to a large and diverse dataset to perform effectively. 
Recently, researchers have explored the potential of in-context learning (ICL) to overcome this challenge in TTS. By employing a target speech prompting strategy, ICL enables systems to generate speech for previously unseen speakers without the need for a pretrained speaker encoder \cite{DBLP:conf/icml/QianZCYH19, DBLP:conf/icassp/XiaoZL22}. This innovative approach significantly enhances zero-shot performance by bypassing the requirement for explicit speaker embeddings, making TTS systems more flexible and scalable. 

In this paper, we present a novel zero-shot VC system that combines disentangled prosody control with the ICL capabilities of advanced pretrained models.  Our proposed system aims to achieve high speaker similarity and preservation of prosody in voice conversion. Specifically, (1) we achieve explicit prosody disentanglement from other speech attributes (content and timbre) by proposing a Prosody-Aware Codec Encoder (PACE), enabling finer control over expressive variations. (2) We leverage the pretrained VALL-E X model \cite{zhang2023speak}, a well-performed TTS system with emergent ICL capabilities, as the backbone of our VC system. This allows for high-quality speech generation while preserving key speaker attributes, even for unseen speakers. (3) To align PACE-generated audio codes with those of VALL-E X, we train the PACE module using VALL-E X audio codes as targets. Our results show that our VC system outperforms baselines in speech quality, timbre similarity, and prosody controllability, enabling a high-quality zero-shot VC system that preserves both speaker identity and prosodic consistency.

%% file: sections/method.tex
\subsection{Overall Architecture}\label{overall_arch}

\begin{figure}[t]
  \centering
  \includegraphics[width=0.9\linewidth]{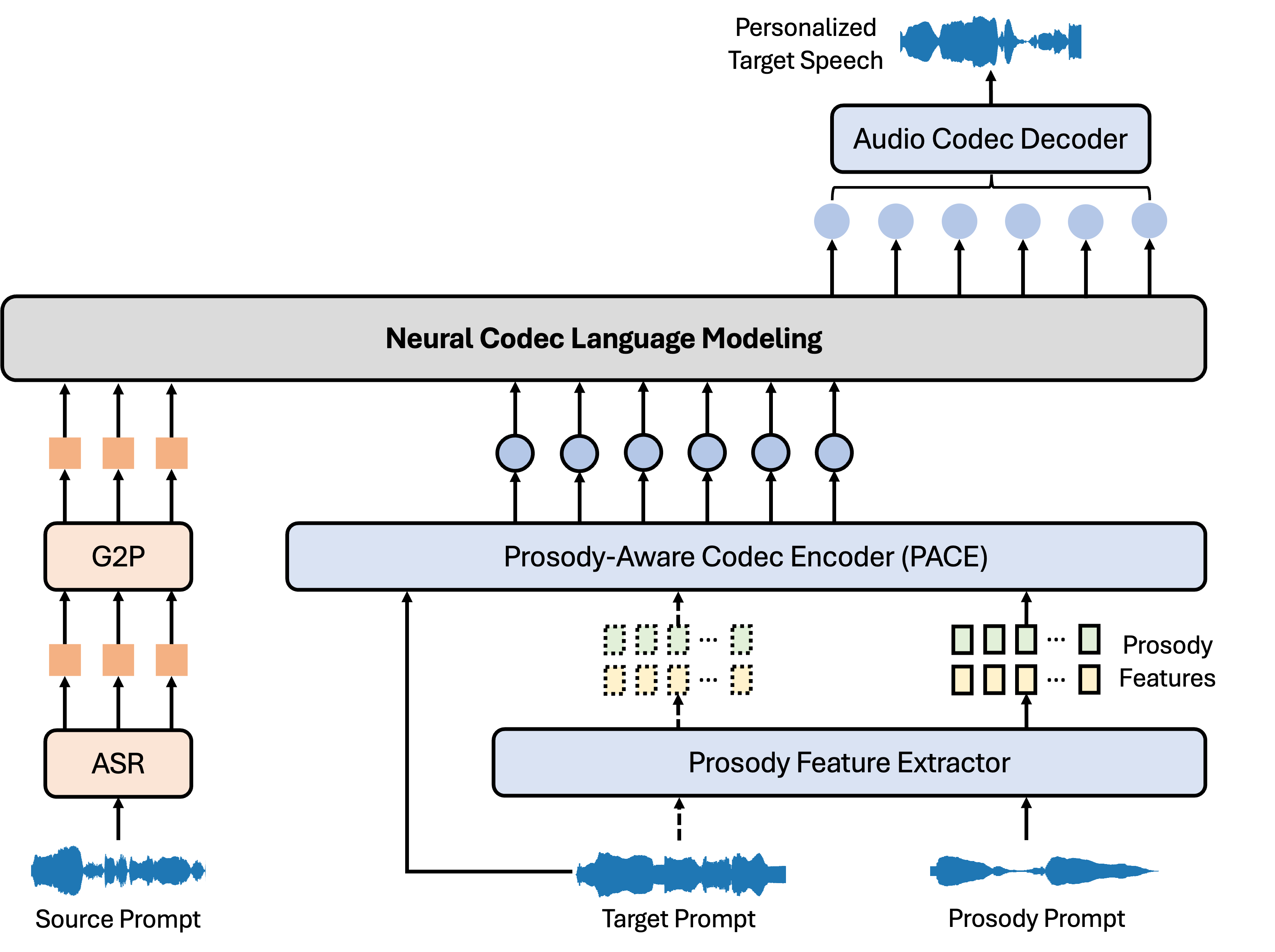}
  \caption{Overall architecture of the proposed voice conversion system. Dotted lines denote components used only during training. Prosody features are derived from the prosody prompt during inference.}
  \label{fig:vc}
\end{figure}

The proposed voice conversion (VC) system, depicted in Figure \ref{fig:vc}, is built upon VALL-E X \cite{zhang2023speak}, a state-of-the-art (SOTA) end-to-end text-to-speech (TTS) model. VALL-E X has demonstrated strong generalization across diverse languages and speech tasks, particularly in zero-shot cross-lingual TTS and speech-to-speech translation (S2ST). The original VALL-E X framework processes both a text prompt and a speech prompt as inputs. The text prompt is transcribed into a phoneme sequence using a grapheme-to-phoneme (G2P) module, while the audio codec encoder \cite{DBLP:journals/taslp/ZeghidourLOST22} maps the speech prompt to an embedding representation. This embedding undergoes residual vector quantization (RVQ) to derive the corresponding audio codes. The neural codec language model then autoregressively predicts audio codes conditioned on the phoneme sequence and speech codes, and the audio codec decoder subsequently synthesizes the output speech waveform.

To exploit the emergent in-context learning (ICL) capabilities of VALL-E X in timbre modeling for VC, we adapt the model for this task. In the VC setting, the model takes as input a source speech prompt \(\mathbf{w}^s \in \mathbb{R}^{L_s \times 1}\) and a target prompt \(\mathbf{w}^t \in \mathbb{R}^{L_t \times 1}\). The objective is to generate the converted speech \(\mathbf{o} \in \mathbb{R}^{L_o \times 1}\), where \(L_s\), \(L_t\), and \(L_o\) denote the lengths of the source prompt, target prompt, and output, respectively, and the second dimension $1$ indicates monaural audio. The generated speech retains the linguistic content of the source speech prompt while adopting the style of the target prompt.


For the source speech prompt, we use Whisper-Medium \cite{DBLP:conf/icml/RadfordKXBMS23} for automatic speech recognition (ASR) to transcribe the speech into text, which is then processed by a grapheme-to-phoneme (G2P) module to obtain the phoneme sequence $\mathbf{p} \in \{0, \dots, N-1\}^{S}$, where $S$ is the sequence length and $N$ is the phoneme vocabulary size.

According to the target speech prompt, a straightforward approach is to directly input it into the audio codec encoder to obtain speech codes. However, this approach lacks prosody control. To address this, we extract prosodic features from the target prompt, including fundamental frequency (\(f_0\)) and unvoiced/voiced (\(uv\)) indicators, following \cite{DBLP:journals/corr/abs-2309-03364, DBLP:journals/cssp/PamisettyM23, DBLP:journals/nn/AnSYX21}. We further introduce the Pitch-Aware Codec Encoder (PACE) module, which derives speech codes conditioned on these prosody features. The PACE module is trained using the extracted prosody features alongside the target speech prompt. 

\subsection{Prosody-Aware Codec Encoder (PACE)}\label{PACE}
\begin{figure}[t]
  \centering
  \includegraphics[width=1\linewidth]{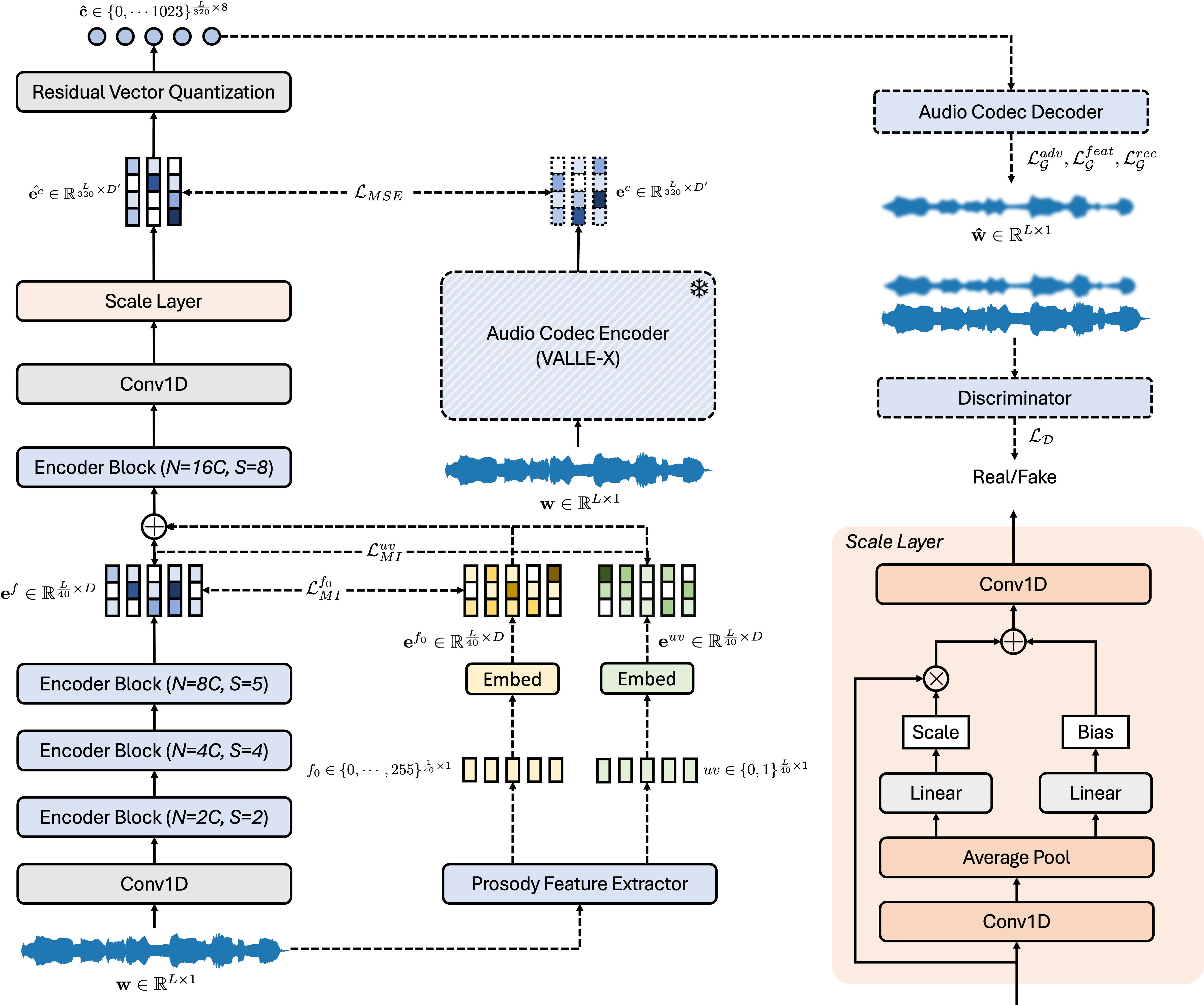}
  \caption{Architecture of proposed Prosody-Aware Codec Encoder (PACE) module. Dotted lines denote components used only during training. Prosody features and embeddings are derived from the prosody prompt during inference.}
  \label{fig:pace}
\end{figure}

The Prosody-Aware Codec Encoder (PACE) module, as shown in Figure \ref{fig:pace}, is designed to extract the audio codec representation from the target speech prompt while conditioning on prosody features. PACE is built upon the audio codec encoder from SoundStream \cite{DBLP:journals/taslp/ZeghidourLOST22} but is structurally modified by splitting the encoder into two stages. The original codec encoder consists of four convolutional blocks, downsampling speech length by factors of $[2, 4, 5, 8]$; we retain the first three layers to extract the speech embedding $\mathbf{e}^f \in \mathbb{R}^{\frac{L}{40} \times D}$, where $D$ denotes the embedding dimension, and $\frac{L}{40}$ represents the sequence length after downsampling.

To disentangle the prosody information from the $\mathbf{e}^f$, we minimize the mutual information (MI) estimation between $\mathbf{e}^f$ and the prosody embeddings $\mathbf{e}^{f_0}$, $\mathbf{e}^{uv}$ through contrastive log-ratio upper bound (CLUB), following \cite{DBLP:journals/tmm/ChenXCZTH23, DBLP:conf/icassp/HuangYLCWXZK23, DBLP:conf/interspeech/YangTZ0SWCTZWM22}. We first extract the prosody features ($f_0$, $uv$) via a Prosody Feature Extractor. Specifically, we employ the harvest function from pyworld\footnote{https://github.com/JeremyCCHsu/Python-Wrapper-for-World-Vocoder} to compute $f_0$ and $uv$ with a frame shift of $\frac{40}{24000} \times 1000 = \frac{5}{3}$ ms, ensuring length alignment with $\mathbf{e}^f$. The $f_0$ sequence is then normalized to $[0,1]$, quantized into 256 discrete values, and represented as $f_0 \in \{0, 1,\cdots, 255\}^{\frac{L}{40}}$, while uv is as $uv \in \{0,1\}^{\frac{L}{40}}$. The MI minization estimation for both $f_0$ and $uv$ is shown in Equation \ref{eq:f0} and \ref{eq:uv}.
\begin{equation}
    \begin{aligned}
    &\mathcal{L}_{MI}^{f_0} = \mathbb{E}_{p\left(\mathbf{e}^{f_0}, \mathbf{e}^f\right)}\left[\log q\left(\mathbf{e}^{f_0} \mid \mathbf{e}^f\right)\right] \\
    &\quad - \mathbb{E}_{p\left(\mathbf{e}^{f_0}\right) p\left(\mathbf{e}^f\right)}\left[\log q\left(\mathbf{e}^{f_0} \mid \mathbf{e}^f\right)\right] \\
    &= \frac{1}{N} \sum_{i=1}^N\left[\log q_\theta\left(\mathbf{e}^{f_0}_i \mid \mathbf{e}^f_i\right)- \frac{1}{M} \sum_{j=1}^M\log q_\theta\left(\mathbf{e}^{f_0}_j \mid \mathbf{e}^f_i\right)\right],
    \end{aligned}
    \label{eq:f0}
\end{equation}
\begin{equation}
    \begin{aligned}
    &\mathcal{L}_{MI}^{uv} = \\
    &\frac{1}{N} \sum_{i=1}^N\left[\log q_\theta\left(\mathbf{e}^{uv}_i \mid \mathbf{e}^f_i\right)- \frac{1}{M} \sum_{j=1}^M \log q_\theta\left(\mathbf{e}^{uv}_j \mid \mathbf{e}^f_i\right)\right].
    \end{aligned}
    \label{eq:uv}
\end{equation}
After obtaining the prosody-invariant audio embedding $\mathbf{e}^f$, we incorporate the prosody information back by summing it with the prosody embeddings $\mathbf{e}^{f0}$ and $\mathbf{e}^{uv}$. These embeddings are derived by passing $f_0$ and $uv$ through an embedding layer.

To ensure the generated audio codes align with those used in the neural codec language model, we employ the trained codec encoder from VALLE-X to first generate the target audio embedding $\mathbf{e}^c \in \mathbb{R}^{\frac{L}{320} \times D'}$, which serves as the input to the residual vector quantizer (RVQ), where $D'$ denotes the embedding dimension, and \( \frac{L}{320} \) results from downsampling \( \frac{L}{40} \) by a factor of 8. We propose a scale layer to align the values of the predicted $\hat{\mathbf{e}^c}$ and target $\mathbf{e}^c$ embeddings within a similar range. This scale layer consists of a Conv1D layer and an Average Pooling layer, which predict a scaling factor $K$ and a bias term $B$. These values are then applied to scale the audio codec embedding, which is subsequently passed through another Conv1D layer, as illustrated in Equation \ref{eq:scale_layer} to obtain the scaled audio codec embedding $\hat{\mathbf{e}^c}$. 
\begin{equation}
    \begin{aligned}
       \hat{\mathbf{e}^c} = \mathrm{Conv1D}(K \times \tilde{\mathbf{e}^c} + B), 
    \end{aligned}
    \label{eq:scale_layer}
\end{equation}
where $\tilde{\mathbf{e}^c}$ denotes the audio codec embedding before scaling. 
To ensure accurate reconstruction, we minimize the mean squared error (MSE) loss between the predicted and target audio codec embeddings as defined Equation \ref{eq:mse}, 
\begin{equation}
    \begin{aligned}
       \mathcal{L}_{recon}^e &= \mathcal{L}_{MSE}(\hat{\mathbf{e}^c}, \mathbf{e}^c).
    \end{aligned}
    \label{eq:mse}
\end{equation}
The generated audio codes $\mathbf{\hat{c}} \in \{0, 1, \cdots, 1023\}^{\frac{L}{320} \times 8}$, where 8 represents the number of codebooks used, are then obtained by passing the predicted audio embedding $\hat{\mathbf{e}^c} \in \mathbb{R}^{\frac{L}{320} \times D'}$ through the RVQ, conditioned on the prosody features.

Beyond the mutual information loss, we adopt the the adversarial loss $\mathcal{L}_{adv}$, the feature loss $\mathcal{L}_{feat}$, and the multi-scale spectral reconstruction loss for generator, $\mathcal{L}_{rec}$ following \cite{DBLP:journals/taslp/ZeghidourLOST22}. These losses jointly guide the training of the PACE module, facilitating high-quality speech generation. Additionally, we follow \cite{DBLP:journals/taslp/ZeghidourLOST22} in employing a discriminator $\mathcal{D}$ with the corresponding loss $\mathcal{L}^\mathcal{D}$ to enhance perceptual quality. The overall objective for generator loss of the PACE module is formulated in Equation \ref{eq:total}, 
\begin{equation}
    \begin{aligned}
       \mathcal{L}_{\mathcal{G}} &= \lambda_{MI}\mathcal{L}_{MI} + \lambda_{recon}^e\mathcal{L}_{recon}^e + \lambda_{adv}\mathcal{L}_{adv} \\
       &\quad + \lambda_{feat}\mathcal{L}_{feat} + \lambda_{rec}\mathcal{L}_{rec}, 
    \end{aligned}
    \label{eq:total}
\end{equation}
where $\mathcal{L}_{MI} = \mathcal{L}_{MI}^{f_0} + \mathcal{L}_{MI}^{uv}$.

\subsection{Training and Inference Scheme}\label{train_inf}
The model training follows a three-stage process. In the first stage, the PACE module is trained without $f0$ and $uv$ as input by minimizing the loss of reconstruction of the audio codec embedding $\mathcal{L}^e_{rec}$. In the second stage, the prosody information is disentangled using the codec encoder parameters learned in stage one, optimizing mutual information loss $\mathcal{L}_{MI}$ along with $\mathcal{L}^e_{rec}$. Finally, in the third stage, the PACE encoder, audio codec decoder are jointly trained by optimizing the total loss.

During the inference phrase, we first extract the phoneme sequence $\mathbf{p}$ from the source prompt $\mathbf{w^s}$ with speaker A. Next, we obtain the prosody features ($f_0$, $uv$) from the prosody prompt $\mathbf{w^p}$ (which can be any speech prompt) using the Prosody Feature Extractor, with speaker X (can be A and B). These prosody features, along with the target prompt $\mathbf{w^t}$ (from speaker B), are then input into the PACE module to generate the prosody-adaptable speech codes $\mathbf{\hat{c}}$. Subsequently, the neural codec language model is utilized to generate the target speech codes $\mathbf{c^o}$, which are passed through the audio codec decoder to produce the target speech $\mathbf{o}$. This generated speech $\mathbf{o}$ retains the content of $\mathbf{w^s}$, the timbre of $\mathbf{w^t}$, and the prosody of $\mathbf{w^p}$. 

%% file: sections/results.tex
\subsection{Training Setup}\label{subsec:train_set}
We train our model on a 54-hour LibriTTS-clean-100 dataset \cite{DBLP:conf/interspeech/ZenDCZWJCW19} and evaluate it on the test-clean set. The training set consists of 33,236 speech samples from 247 speakers, all resampled to 24 kHz. For zero-shot evaluation, we select 10 male and 10 female speakers from LibriTTS. During training, we randomly extract 2-second segments from the speech clips, with zero-padding applied to clips shorter than 2 seconds. 


The encoder and decoder of PACE module follows the architecture in \cite{DBLP:journals/taslp/ZeghidourLOST22}. The \( f_0 \) and \( uv \) embedding layers have a dimension of 256, with vocabulary sizes of 256 and 2. The scale layer includes a Conv1D layer \((128, 64)\) with a kernel size of 3 for scale and bias extraction, followed by a linear layer \((64, 1)\), and a final Conv1D layer \((128, 128)\) with a kernel size of 3. Our experiments are implemented in PyTorch and PyTorch Lightning, our model was trained on NVIDIA RTX A5000 GPUs for 360k, 60k, and 180k steps across the three training stages.

\subsection{Evaluation Methods}\label{subsec:eval_methods}
For evaluation, we use the following metrics. The ASV-Score \cite{zhang2023speak} measures speaker similarity by calculating the cosine distance between speaker embeddings from a pretrained WavLM model. The ASR-WER \cite{DBLP:conf/icml/RadfordKXBMS23} evaluates intelligibility by comparing synthesized speech to ground truth transcriptions using the Whisper model. We assess naturalness with the NISQA-TTS model\footnote{https://github.com/gabrielmittag/NISQA} \cite{DBLP:conf/interspeech/MittagNC021}, providing a score from 0 to 5 for fluency, clarity, and expressiveness. The MOSNET model\footnote{https://github.com/aliutkus/speechmetrics} \cite{DBLP:conf/interspeech/LoFHWYTW19} offers an objective MOS score, reflecting overall speech quality. For prosody matching, we compute the F0 distance \cite{DBLP:conf/interspeech/LuWL0Z20}, measured as the normalized distance between the F0 contours of converted and reference speech. Subjective evaluation involved mean opinion scores (MOS) and speaker similarity (SMOS), with 22 participants rating audio samples from various systems on a scale of 1 to 5, where 5 represented the highest quality.

\subsection{Effectiveness of Pitch-Aware Codec Encoder (PACE) module}
\begin{table}[t]
\caption{Evaluation for PACE v.s. Baseline encodec encoder.}
\label{tab:encodec}
\begin{center}
\resizebox{\columnwidth}{!}{
\begin{tabular}{l|cccc}
\toprule
\textbf{Model} & \textbf{ASV (↑)} & \textbf{ASR-WER (↓)} & \textbf{NISQA (↑)} & \textbf{MOSNET (↑)} \\
\midrule
Baseline Codec Encoder & 0.6813 & 0.1239 & 4.1662 & 3.8755 \\
PACE & 0.6620 & 0.1104 & 3.9805 & 3.6592 \\
\midrule
Ground Truth & - & - & 4.6382 & 4.0233\\
\bottomrule
\end{tabular}
}
\end{center}
\end{table}
We compare PACE module with the pretrained 24 kHz EnCodec\footnote{https://huggingface.co/facebook/encodec\_24khz}, the audio codec encoder originally used in VALL-E X (Baseline Codec Encoder), across objective evaluation metrics. As shown in Table \ref{tab:encodec}, while there remains a slight gap in performance between the original EnCodec and our proposed PACE module, this difference does not significantly impact overall performance. This suggests that our PACE module effectively retains much of the original EnCodec's capabilities. Notably, the ASR-WER demonstrates a substantial improvement, reflecting enhanced speech intelligibility.

\subsection{Voice Conversion: Speaker Timbre Control and Overall Quality}\label{subsec:vc_overall}
\begin{table}[t]
\caption{Objective and Subjective Evaluation of Proposed and Baseline VC Systems.}
\label{tab:overall}
\begin{center}
\resizebox{\columnwidth}{!}{
\begin{tabular}{l|cccc|cc}
\toprule
\textbf{Model} & \textbf{ASV (↑)} & \textbf{ASR-WER (↓)} & \textbf{NISQA (↑)} & \textbf{MOSNET (↑)} & \textbf{MOS (↑)} & \textbf{SMOS (↑)}\\
\midrule
VALLE-X & 0.8429 & 0.1151 & 4.3034 & 3.6059 & 4.1880 & 3.8251\\
TriAAN-VC & 0.7199 & 0.1298 & 4.2527 & 3.5451 & 4.0667 & 3.7870\\
Proso-VC & 0.7025 & 0.1632 & 3.2520  & 2.9105 & 3.9520  & 3.2542\\
\midrule
Ours & 0.9078 & 0.1010 & 4.3320 & 3.5746 & 4.3627 & 3.9386\\
-w/o $\mathcal{L}_{MI}$ & 0.8751 & 0.1314 & 4.2892 & 3.4178 & 3.7696 & 3.4074\\
-w/o scale layer & 0.8771 & 0.1574 & 3.9424 & 2.7236 & 3.2876 & 2.8333\\
-w/o $\mathcal{L}_{rec}^e$ & 0.5882 & 0.2286 & 3.0974 & 2.0822 & 2.4831 & 2.3550\\
\midrule
Ground Truth & - & - & 4.5890 & 3.9328 & 4.3940 & 4.1585\\
\bottomrule
\end{tabular}
}
\end{center}
\end{table}
We evaluated the overall quality of voice conversion by comparing our model with baseline models: VALLE-X, TriAAN-VC, and ProsoVC, as shown in Table \ref{tab:overall}. TriAAN-VC \cite{DBLP:conf/icassp/ParkYKSH23} enables nonparallel conversion from any to any through adaptive attention normalization, while ProsoVC \cite{DBLP:conf/icassp/ZhaoLSWKTM22} controls prosody using hybrid bottleneck features. Our model outperforms these baselines in most metrics, achieving the highest ASV score, a lower ASR-WER, and the best NISQA score, indicating superior preservation, intelligibility, and naturalness of the speaker identity. Although the MOSNET score is slightly lower than the VALLE-X, it is still higher than other baselines, underscoring our model's overall effectiveness in voice conversion. Ablation studies demonstrate the effectiveness of our method, significantly outperforming the model without the scale layer and reconstruction loss training, while achieving a slight improvement over the variant without mutual information loss training. Subjective evaluation results indicate that our model achieved strong performance in both naturalness and speaker similarity, particularly outperforming ablated variants, demonstrating the effectiveness of the proposed modules. 

\subsection{Voice Conversion: Capability in Prosody Alignment}\label{subsec:prosody_eval}

\begin{figure}[t]
  \centering
  \includegraphics[width=\linewidth]{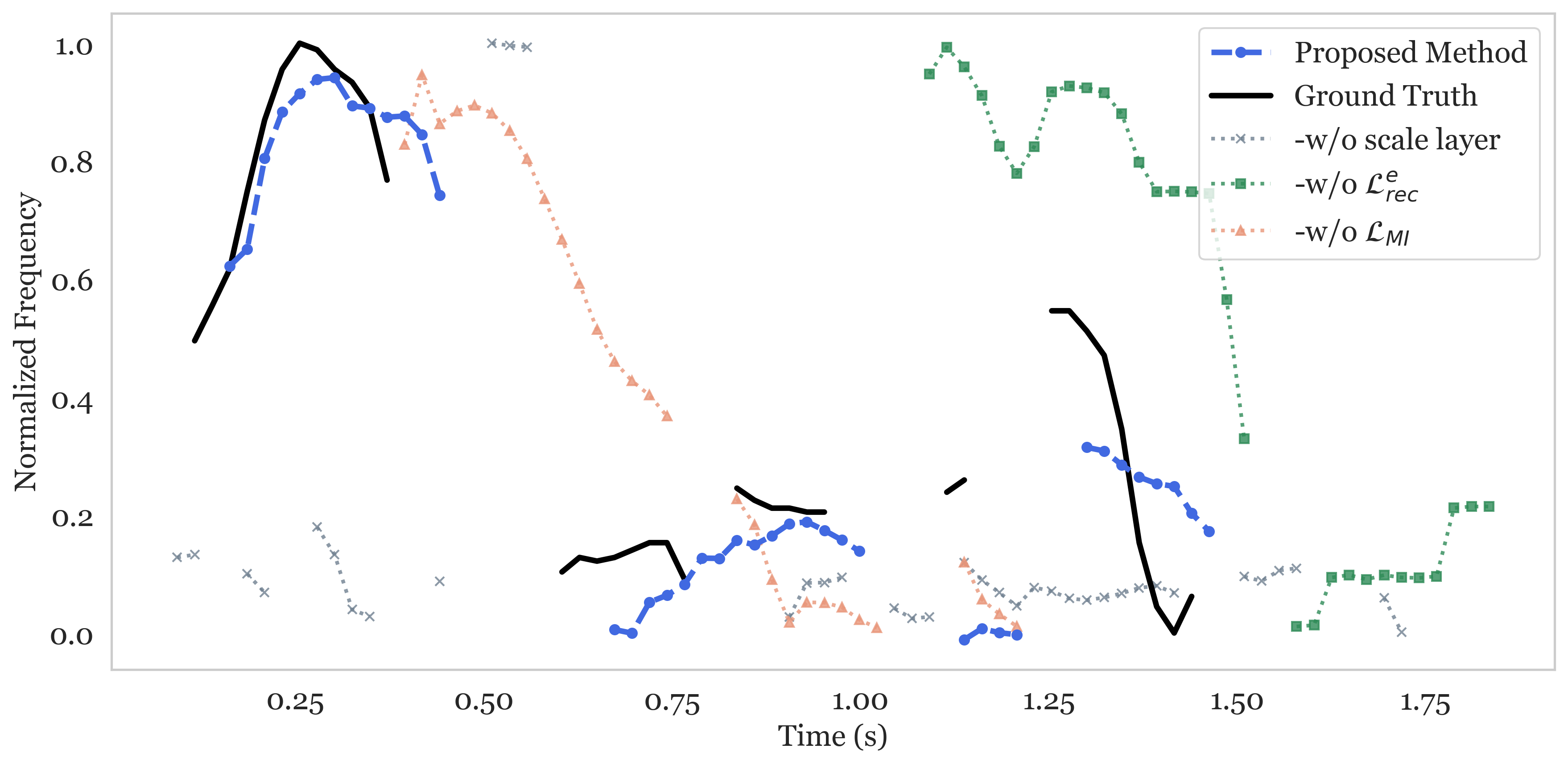}
  \caption{F0 contour comparison for our proposed method and reference prosody prompt.}
  \label{fig:prosody}
\end{figure}

\begin{table}[t]
\caption{Prosody Matching in Voice Conversion (F0-scaled Distance).}
\label{tab:prosody}
\begin{center}
\resizebox{\columnwidth}{!}{
\begin{tabular}{l|cc}
\toprule
\textbf{Models} & \textbf{Prosody from Source} & \textbf{Prosody from Target}\ \\
\midrule
VALL-E X & - & 3.1029 \\
TriAAN-VC & 3.4019 & - \\
ProsoVC & 3.3256 & - \\
\midrule
Ours & 2.8239 & 2.6988\\
-w/o $\mathcal{L}_{MI}$ & 3.2751 & 3.0179\\
-w/o scale layer & 3.9178 & 3.6237\\
-w/o $\mathcal{L}_{rec}^e$ & 4.7892 & 4.2649\\
\bottomrule
\end{tabular}
}
\end{center}
\end{table}

We evaluate prosody matching in two scenarios: prosody from the source speech and from a prompt. Results in Table \ref{tab:prosody} are compared with VALLE-X, TriAAN-VC, and ProsoVC using normalized F0 distance. VALLE-X is excluded from the "source" scenario, and TriAAN-VC/ProsoVC from the "prompt" scenario, as they lack these functions.

Our model outperforms the baselines in both scenarios. In the "prosody from source" case, it leads, while VALLE-X, limited to prompt-based prosody conversion, cannot perform in this scenario. In the "prosody from prompt" case, VALLE-X shows reasonable performance due to the ICL capability but still lags behind our model, emphasizing the effectiveness of our disentanglement approach for prosody matching. We visualize the normalized F0 curves of the synthesized output and the reference prosody prompt. As shown in Figure \ref{fig:prosody}, our method's F0 curve closely aligns with the reference, demonstrating its effective prosody control. In ablation studies, our model outperforms the ablated variants in both scenarios, with a more pronounced advantage in the "Prosody from Source" setting. While the model trained without $\mathcal{L}_{MI}$ achieves comparable performance in overall quality and timbre similarity, the results highlight the importance of mutual information loss in prosody control.

%% file: sections/concl.tex
We propose a voice conversion (VC) model that explicitly disentangles prosody from speaker timbre, enabling precise prosody control while preserving speaker identity. To achieve this, we introduce the prosody-aware audio codec encoder (PACE), which conditions audio codes on specific prosodic features, allowing fine-grained manipulation of prosody. By isolating prosody from other acoustic attributes, our approach enhances prosody control, ensuring greater flexibility in voice conversion tasks. Comprehensive evaluation results demonstrate that our model outperforms baseline systems in prosody alignment, timbre consistency, and overall speech quality. In particular, our method achieves more natural and expressive voice conversion that closely matches the target speaker’s style. These results underscore the effectiveness of our disentanglement strategy and its broader applicability in controllable speech synthesis and expressive voice conversion.